\documentclass[pra,twocolumn,showpacs,floatfix]{revtex4}
\usepackage{graphicx}

\newcommand{\beq}{\begin{equation}}
\newcommand{\eeq}{\end{equation}}
\newcommand{\beqa}{\begin{eqnarray}}
\newcommand{\eeqa}{\end{eqnarray}}
\newcommand{\ba}{\begin{array}}
\newcommand{\ea}{\end{array}}

\begin{document}

\title{Universal behavior of a  trapped Fermi superfluid
 in the BCS-unitarity crossover}
\author{S. K. Adhikari
}
\affiliation{
Instituto de F\'{\i}sica Te\'orica, UNESP - S\~ao Paulo State
University, 01.405-900 S\~ao Paulo, S\~ao Paulo, Brazil}

\begin{abstract}

From an extensive calculation of static properties of a 
trapped Fermi 
superfluid
at zero temperature  using a  density-functional 
formulation, we demonstrate a universal behavior of its observables, 
such as energy, chemical
potential,  radius etc., over 
the crossover from the  BCS  limit  to unitarity 
leading to scaling  over many orders of magnitude in fermion number. 
This scaling  allows  to predict the static properties of the system,
with a 
large  number ($\sim 10^5$) of fermions,  over the crossover with an 
error of 
1-2$\%$, 
from the  knowledge of those for a small  number ($\sim 10$) of 
fermions. 

\end{abstract}

\pacs{03.75.Ss, 71.10.Ay, 67.85.Bc, 05.30.Fk}
\maketitle

\section{Introduction}

The Pauli principle leads to an effective repulsion
in identical fermions which could dominate the  physical
interaction and control the dynamics. 
The dominance of Pauli repulsion is responsible for the stability of our 
hadronic universe. 
When this happens the system exhibits universal behavior practically 
independent of or weakly dependent on the physical
interaction.  One classic 
example \cite{amado} of this  is found in the s-wave quartet 
nucleon-deuteron 
system  with three spin-parallel 
nucleons controlled by  Pauli repulsion. The calculated scattering 
length  for different  nucleon-nucleon interactions is
essentially  6.3 fm, whereas the doublet s-wave 
scattering length (not controlled by Pauli repulsion) for the same 
interactions varies from $-2$ fm to 3 fm \cite{amado,tomio}. 
This 
universality is
 prominent in the limit of zero  Fermi-Fermi
interaction  in the Bardeen-Cooper-Schrieffer (BCS) theory of 
superfluid fermions leading to universal properties of low-temperature 
superconductors \cite{bcs}, of cold neutron matter and neutron star 
\cite{baker,th1}, 
and of 
a  trapped     Fermi superfluid \cite{review} at zero temperature.    
This 
universality also manifests \cite{castin,blume,blume2,blume3} in a 
trapped  Fermi 
superfluid  
at  unitarity as the Fermi-Fermi scattering length $a$ goes 
to 
infinity ($a\to -\infty$). In this limit, though the physical 
interaction 
is nonzero, the only interaction  scale (scattering length) 
disappears 
and the system acquires universal behavior. This 
interaction scale is also absent in the  BCS limit ($a=-0$) with 
universal properties. Similar universality is found in other fermionic 
systems with large $|a|$ \cite{braaten}. 

The crossover from   weak-coupling 
BCS
limit to unitarity  
 \cite{1}
has been a very  active area of 
research  \cite{th1,th3,blume,castin}  after 
the experimental realization  \cite{excross,excross1} 
of  this  crossover 
 in a trapped dilute Fermi superfluid
near a Feshbach resonance. Using a complete numerical simulation of a 
density-functional (DF) formulation \cite{df}, we show that the 
deviation from 
universality of a trapped dilute Fermi superfluid  over the crossover 
is  orderly
and the system continues to exhibit nearly universal 
behavior possessing useful scaling relations over many 
orders of magnitude involving energy, chemical potential, radius and  
number of Fermi atoms ($N$). In this 
crossover region the    Pauli repulsion  dominates over the 
physical interaction 
leading to  the universal behavior.

In Sec. II we present the density functional formulation that we use 
in this study. In Sec. III we present the numerical results and establish 
the universal behavior of a trapped Fermi superfluid in the BCS-unitarity crossover.
Finally, in Sec. IV  we give some concluding  remarks.

\section{Density Functional Formulation}

To study the universality, we use a 
Galilei-invariant DF formulation for the crossover of a trapped 
two-component Fermi 
superfluid \cite{LS}, equivalent to a
hydrodynamical model with  the correct 
phase-velocity relation \cite{review} ${\bf v}=\hbar \nabla \theta 
/(2m)$, where ${\bf 
v}$ is the superfluid velocity, $m$ the Fermi mass, and $\theta$ the 
phase of the order parameter $\Psi({\bf r})$ at position ${\bf r}$, 
which satisfies (Eq. (35) of \cite{LS}, but with a distinct  
$g(x)$ consistent with the known small-$x$  behavior of energy of a 
uniform Fermi gas)
\begin{eqnarray}\label{1}
&&\left[-\frac{\hbar^2}{8m}\nabla^2+U+ 
\mu(n,a)\right]\Psi({\bf r})=\mu_0\Psi({\bf r}),\\
\label{2} 
&& \mu(n,a)=\frac{\hbar^2}{2m}(3\pi^2n)^{2/3}g({n^{1/3}a}),\\
\label{3}
&&\mu_0=\int d {\bf r}\biggr[\frac{\hbar^2}{8m}|\nabla \Psi|^2+U\Psi^2
+\mu(n,a)\Psi^2\biggr],\label{4}
\end{eqnarray}
where $\int \Psi^2({\bf r}) d{\bf r}=N$, and 
\begin{equation}\label{gnew}
g(x)=1+\frac{(\chi_1 x-\chi_2 x^2)}{(1-\beta_1 x+\beta_2 x^2)},
\end{equation} 
with
$\chi_1 = 4\pi/(3\pi^2)^{2/3}, \chi_2=300, \beta_1=40,$ 
$\beta_2=\chi_2/(1-\zeta)$,  $\mu(n,a)$  
the bulk chemical 
potential, $\mu_0$ the chemical potential for the 
trapped system,  
$n=\Psi^2$  the density of  atoms, 
and $U =m\omega^2 r^2/2$ 
the harmonic trap of frequency $\omega$. 
(Different parametrizations of $g(x)$ were used in Ref. \cite{LS1}.)
Here  
we take 
$\zeta=0.44$  consistent with Monte Carlo calculations 
\cite{th1,th3} and experiments \cite{exp} of a uniform Fermi superfluid  
at unitarity. 
The  
parameters $\chi_1,\chi_2,\beta_1$ and $\beta_2$  are chosen so that the 
model 
(i) agrees with the fixed-node Monte Carlo (FNMC) \cite{blume}  and 
Green-function Monte 
Carlo (GFMC) \cite{CB}  results for  the energy 
of a trapped  superfluid 
at   unitarity (for $N<30$)
and over the crossover\cite{LS,blume},  (ii) 
provides a 
smooth 
interpolation between the energies of a  superfluid 
at the BCS and unitarity limits \cite{LS,vs}, and (iii) 
satisfies  the 
known BCS limit \cite{LS3} of 
the bulk chemical potential $\mu(n,a)$ (two lowest-order terms of Eq. 
(1) of 
\cite{LS2}). 

\begin{figure}[tbp]
\begin{center}
{\includegraphics[width=1\linewidth]{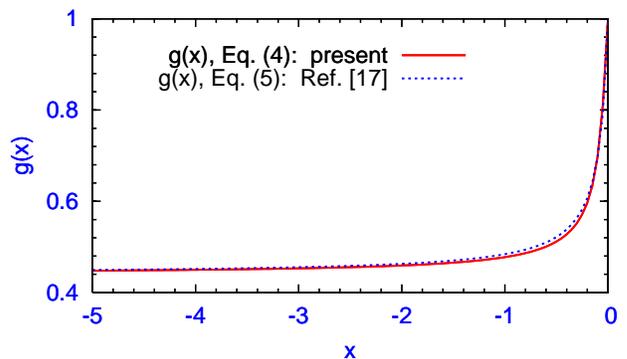}}
\end{center}
\caption{(Color online) 
The function $g(x)$ vs. $x$ as defined by Eqs. (\ref{gnew}) and 
(\ref{gold}).
}
\label{fig0}
\end{figure}

In Ref. \cite{LS} we used the following simple expresion for the 
function $g(x)$ in place of that given by Eq. (\ref{gnew}):
\begin{equation}\label{gold}
g(x)= 1+\frac{\chi x}{1-\beta x}, 
\end{equation}
with $\chi=20 \pi /(3 \pi^2)^{2/3}$, and $\beta = \chi/(1-\zeta)$.
This choice does not satisfy the known weak interaction BCS limit 
$\lim_{x\to 0}g(x)\to  1+4 \pi x/(3 \pi^2)^{2/3}$
\cite{LS,LS1,LS2}. Equation (\ref{gold}) was used in Ref. \cite{LS} as a 
simple 
model to 
satisfy 
the   BCS-unitarity crossover Monte Carlo results \cite{vs} for 
energies of trapped fermions.
  The bulk energy (and chemical potential) of a Fermi
   gas in the weak-couping BCS limit can be written as an expansion in
   the parameter $x=n^{1/3}a$ as discussed in Ref. \cite{LS2}. Lee and 
Yang 
\cite{LS3}
   calculated  the coefficients of the series in a specific model of
   interacting fermions as quoted in Eq. (1) of \cite{LS2}. These 
coefficients
   can be related to the   coefficients in an expansion of $g(x)$ in the
   small $x$ limit. The   $x$-independent constant in $g(x)$ and the 
coefficient
   of the $x$ term    $\chi=4\pi/(3\pi^2)^{2/3}$  in this series are 
reasonably 
model independent. 
 However, coefficients of $x^2$ will be dependent on the
   interaction model. In  Ref. \cite{LS} we changed the coefficient of 
the 
$x$-term
   from that proposed in \cite{LS2}, to simulate the effect of the 
(unknown)
  higher order terms in the  $x$-expansion of $g(x)$. The    coefficient 
of
  the $x$-term can be kept at   the Lee-Yang value    
$4\pi/(3\pi^2)^{2/3}$
   provided we include a $x^2$ term in the expansion of $g(x)$ and we 
have
  done this in Eq. (\ref{gnew}). However, the coefficient of the
   present $x^2$ term does not and should not agree with that of the 
known
  expansion quoted   in Eq. (1)
   of \cite{LS2}. This is    because this coefficient now simulates the
   contribution of higher    order terms.  Although the analytic 
expression  $g(x)$ 
is modified here from that given by Eq. (\ref{gold}) to (\ref{gnew}), 
from a numerical (calculational) point of view the change is negligible 
as we can see from Fig. \ref{fig0}, where we plot the two 
parametrizations of the function $g(x) $ given by Eqs. (\ref{gnew}) and 
(\ref{gold}).   In Ref. \cite{LS}, we used the function $g(x)$ of Eq. (\ref{gold}) to 
reproduce some Monte-Carlo results \cite{vs} for energy of 
a trapped-fermion system 
over the BCS to unitarity crossover. We have now performed the same 
calculations
with the  function $g(x)$ of Eq. (\ref{gnew}) $-$ illustrated below in 
Fig. \ref{fig3} (b) $-$ 
and we verified that the results for energy remain practically 
unchanged.

The gradient term in Eq. (\ref{1}) 
provides a correction to 
the 
local 
density approximation (LDA)  \cite{bulgac} obtained by setting the 
gradient term to 
zero.  LDA is a good approximation for a large $N$, when the 
bulk chemical potential $\mu(n,a)$, a positive term responsible for 
Pauli repulsion in 
the system even for attractive (negative) $a$,  
is 
very large. The gradient term is consistent with the hydrodynamic flow 
of paired fermions of mass $2m$ \cite{LS,review}. To study the scaling  
of the 
solution of Eq. (\ref{1}) we note 
that  in the BCS  and unitarity  limits the bulk 
chemical potential $\mu(n,a)$
has, respectively,  the following simple forms 
\cite{review,th3,th1}:
$\hbar^2 (3\pi^2n)^{2/3}/(2m)$ and $\zeta \hbar^2 
(3\pi^2n)^{2/3}/(2m)$, which shows the scaling  $\mu(n,a)\sim 
n^{2/3}$ in both limits.  In these limits the energy functional is given 
by \cite{review}
\beqa
E&=&\int d {\bf r}\biggr[\frac{\hbar^2}{8m}|\nabla \Psi|^2+U\Psi^2
+{3\hbar^2\xi  \over 20m}(3\pi^2 )^{2/3}\Psi^{10/3}\biggr],\label{5} \\
&\equiv& \langle E_{\mathrm{\nabla}} \rangle + \langle E_ {\mathrm{pot}} 
\rangle+ 
\langle E_{\mathrm{Fermi}} \rangle, \label{6}
\eeqa
where $\xi =1$ in the BCS limit and $\xi=\zeta=0.44$ at unitarity 
\cite{th1,th3}, and 
$\langle E_{\mathrm{\nabla}} \rangle$, $\langle E_{\mathrm{pot}} 
\rangle$, 
and 
$\langle E_{\mathrm{Fermi}}\rangle$  are, respectively, the expectation 
values of 
the three terms in Eq. (\ref{5}).
This ``analytic" dependence of $E$ on $\Psi$  
leads to a simple virial theorem, postulated and studied 
experimentally in \cite{z},  
connecting $\langle E_{\mathrm{\nabla}} 
\rangle$, $\langle E_{\mathrm{pot}} \rangle$, and
$\langle E_{\mathrm{Fermi}}\rangle$  at the BCS and unitarity limits. 
For 
the exact $\Psi$, energy $E$ is a minimum. 
To 
derive the virial theorem, we take the Cartesian system ${\bf 
r}\equiv (x,y,z)$, and the norm-preserving scaling transformation 
$\Psi(x,y,z)\to \sqrt \lambda \psi(\lambda x,y,z)$. The condition of 
minimum energy is 
$|d  E  /d \lambda|_{\lambda =1}=0$. The 
$\lambda$-dependent part of energy now becomes $ E_\lambda 
=\lambda^2 \langle E_{\mathrm{\nabla}}^x  \rangle + \lambda^{-2} \langle 
E_{\mathrm{pot}}^x  \rangle +\lambda^{2/3} \langle E{_\mathrm{Fermi}}  
\rangle, $
where the suffix $x$ denotes $x$ component. The minimization condition 
yields $\langle
E_{\mathrm{pot}}^x  \rangle=\langle
E_{\mathrm{\nabla}}^x  \rangle+\langle
E_{\mathrm{Fermi}}  \rangle/3$. By summing over three components, we 
have the 
virial theorem $\langle
E_{\mathrm{pot}}  \rangle=\langle
E_{\mathrm{\nabla}}  \rangle+\langle
E_{\mathrm{Fermi}}  \rangle$, or equivalently, $E=2 
\langle E_{\mathrm{pot}}  
\rangle$  at the BCS and unitarity limits. The deviation in 
percentage from the virial 
theorem in the crossover can be estimated  by  the 
percentage defect function 
$\eta=100 (E-2\langle E_{\mathrm{pot}}\rangle)/E$.

In the LDA  one has 
the following analytic solutions for the energy, chemical 
potential, and mean square radius of the 
trapped system \cite{bulgac}: $E/(\hbar\omega)=(3N)^{4/3}\sqrt \xi/4,  
\mu_0/(\hbar\omega)=(3N)^{4/3}\sqrt \xi/3$, $\langle r^2 
\rangle/({\hbar/m\omega})= 
(3)^{4/3}N^{1/3}\sqrt \xi/4$.  
The absence of the parameter $a$ in 
$\mu(n,a)$ of Eq. 
(\ref{2}) in the BCS and unitarity limits leads to the following 
properties: (i) the  scaling $\mu(n,a) \sim 
n^{2/3}$ for the uniform Fermi superfluid, 
(ii) scaling $E/(\hbar\omega), \mu_0/(\hbar\omega)\sim N^{4/3}$  
and (iii) the virial theorem, $E=2\langle E_{\mathrm{pot}} 
\rangle$
for the trapped Fermi superfluid. 
These universal properties lead to predictability of the Fermi 
superfluid in the BCS and unitarity limits, e.g., predicting the 
energy 
for a large number of fermions from the  knowledge of that for a small 
number of fermions obtained by accurate ``exact" calculation. This is
important as exact calculations for large systems are very difficult, if 
not impossible. 

To extend the  above predictability  over the crossover, where 
analytical (LDA) results are not available, 
we  study the deviation  from the universal 
properties of a Fermi superfluid in the crossover 
region. We find that the deviation is orderly, which allows us 
to restore a universal and predictable behavior of the trapped Fermi 
superfluid, so that one can predict the properties of a large system 
from a  knowledge of those for a small system.      

\section{Numerical Results}

We solve Eq. (\ref{1}) by transforming it to time-dependent form by 
replacing $\mu_0$ by a time derivative. We express energy variables in 
units of $\hbar \omega$, length in $\sqrt{\hbar/(m\omega)}$ and time 
in $\omega^{-1}$.
The resultant equation is 
then discretized by the semi-implicit Crank-Nicholson algorithm using a 
typical
space step 0.04 and time step 0.001 
and then solved by imaginary time propagation.   
The chemical 
potential $\mu_0$ is then calculated via Eq. (\ref{4}), energy in the 
BCS and unitarity limits via Eq. (\ref{5}), and  the energy in the 
crossover 
region via a numerically constructed energy functional. 
In Table 
\ref{table1} we display energy per particle in units of
$E_F=(3N)^{1/3}\hbar\omega$ (the Fermi energy of an ideal Fermi gas
at the trap center) or $E/[\hbar\omega(3N)^{4/3}]$
along the 
crossover for different $N$ and $a$. In Table
\ref{table2} we report the respective chemical potentials  
$\mu_0/[\hbar\omega(3N)^{4/3}]$.

\begin{figure}[tbp]
\begin{center}
{\includegraphics[width=1\linewidth]{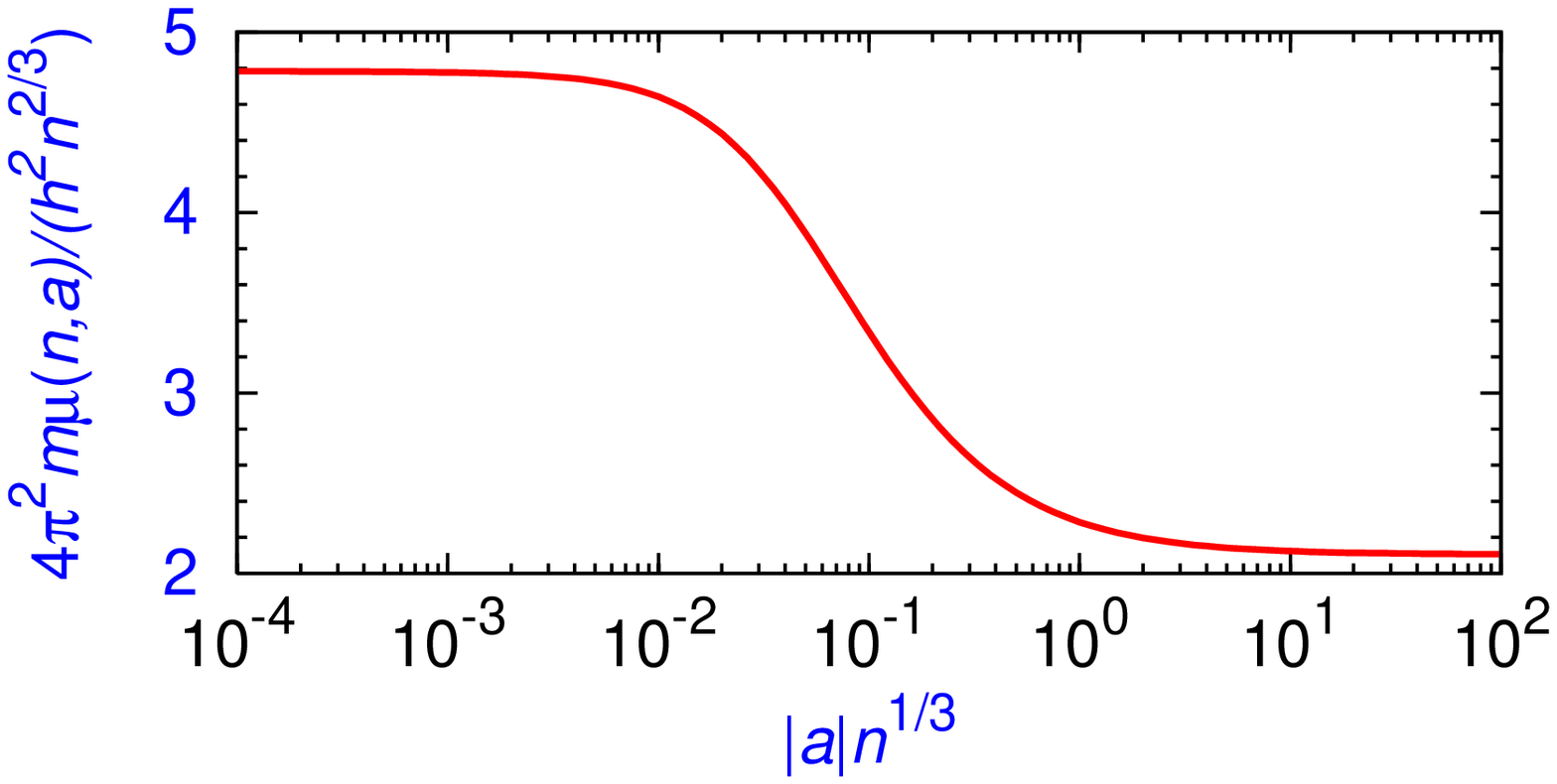}}
{\includegraphics[width=\linewidth]{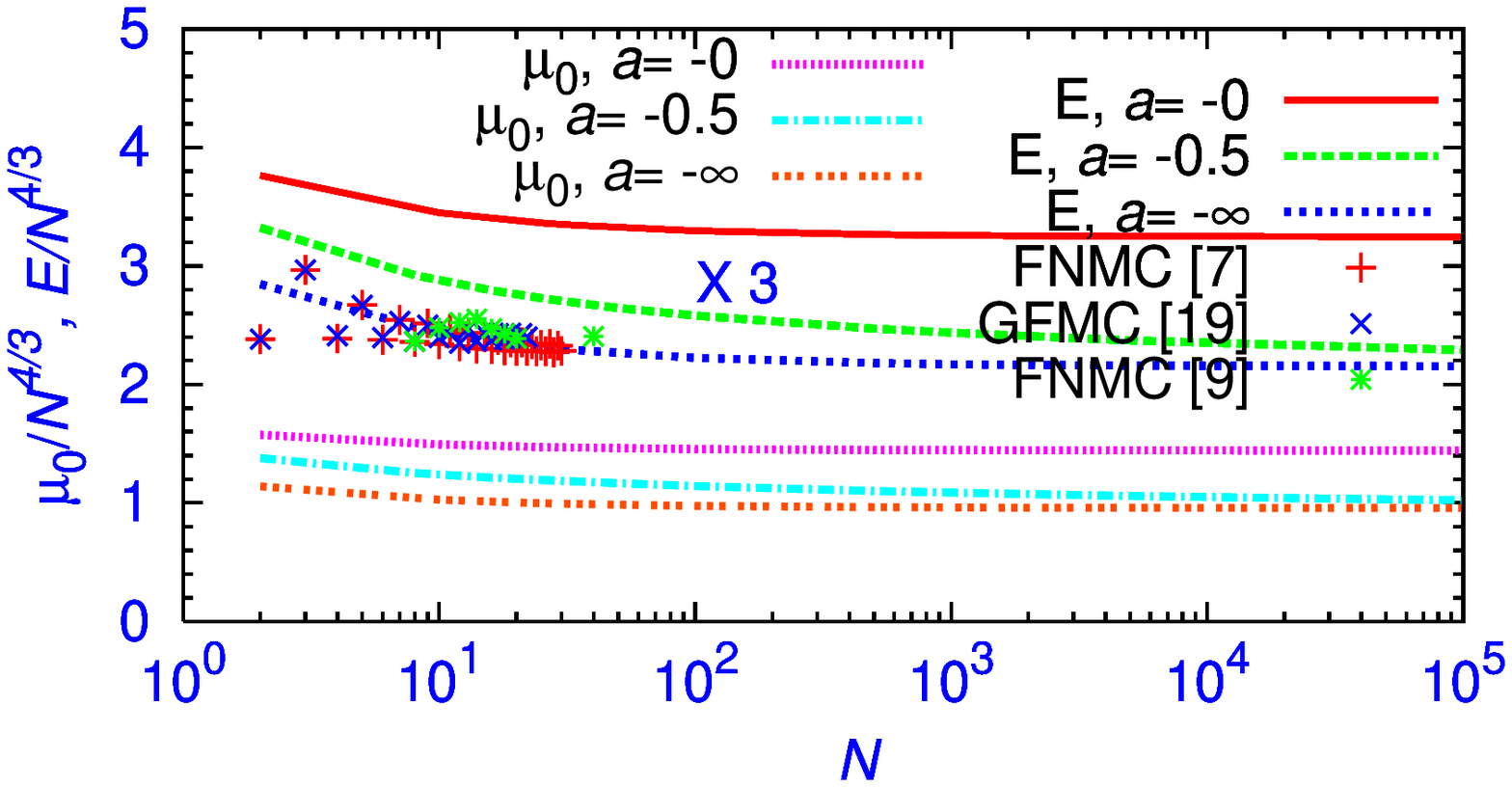}}
\end{center}
\caption{(Color online) (a)Dimensionless bulk chemical potential  
$4\pi^2m\mu(n,a)/(h^2 n^{2/3})$ 
 of a uniform Fermi gas vs. $|a|n^{1/3}$ for  
$a<0$. (b) Chemical potential and energy   
$\mu_0(N,a)/N^{4/3}$ and $E(N,a)/N^{4/3}$ of a trapped Fermi 
gas
vs.  $N$  for  $a=-0, -0.5, -\infty$. 
The energies  of FNMC \cite{blume,blume3} and GFMC \cite{CB} 
calculations 
at unitarity are 
also shown. (Energies and scattering lengths are expressed in oscillator units.)
}
\label{fig1}
\end{figure}

To understand the universal nature of $E$ and $\mu_0$ of Tables 
\ref{table1} and \ref{table2}
for different $N$, we study the nonlinear input $\mu(n,a)$ 
to Eq. (\ref{1}). In Fig. \ref{fig1} (a) we 
plot $4\pi^2 n \mu(n,a)/(h^2n^{2/3})$ vs. $|a|n^{1/3}$  for $a<0$.  In 
the 
BCS ($|a|n^{1/3}\to 0$)
and 
unitarity ($|a|n^{1/3}\to \infty$)
limits one has the perfect scaling, $\mu(n,a) \sim 
n^{2/3}$, with  deviation from this behavior in the crossover   
region. 
Next, to study the behavior of  $E$ and  $\mu_0$  
of the trapped Fermi superfluid, 
we plot 
in Fig. \ref{fig1} (b) the numerically calculated  $E/N^{4/3}$ and   
$\mu_0/N^{4/3}$  vs. $N$ for  $a=-0,-0.5$ and $-\infty$ and the FNMC 
\cite{blume,blume3} and GFMC \cite{CB} results.
Perfect scalings, 
$\mu_0, E \sim N^{4/3}$,  are 
observed for 
large $N$  
in the BCS and unitarity limits,  with deviation in the crossover 
region and  for small $N$.   A careful analysis of the  $\mu_0, E$  
data of Fig.  \ref{fig1} (b)
for small $N$ reveals in the crossover region
the average scalings $(E/N^{2/3}-0.37)\sim 
N^{2/3}$ and 
$(\mu_0/N^{2/3}-0.27)\sim N^{2/3}$.  
If these scalings were really perfect 
then 
plots  of $(E/N^{2/3}-0.37)/
N^{2/3}$  or $(\mu_0/N^{2/3}-0.27)/ N^{2/3}$ vs. $|a|$ for a fixed 
$N$ would lead to universal curves independent of $N$.

\begin{table}[!ht]
\begin{center}
\caption{Dimensionless energies $E/[\hbar\omega(3N)^{4/3}]$ of 
a trapped Fermi 
superfluid along the crossover. 
The last two  columns, \{4\} and \{10\},  give the predicted 
energies for $N=10^5$  employing scaling (\ref{7}) 
using the energies for 
$N'=4$ and 10, respectively.  }
\label{table1}
\begin{tabular}{|r|r|r|r|r|r|r|}
\hline
$a\backslash N=$ & 10  & $10^2$ &
  $10^4$ & $10^5$ &  $10^5$\{4\}&$10^5$\{10\} \\
\hline
   $-0.001$  &0.2656  & 0.2539  & 0.2501  &  0.2499 &0.2428 
&0.2470
\\
   $-0.01$  & 0.2652 & 0.2531  &   0.2478 &  0.2459  & 0.2393  
& 0.2434  \\
   $-0.1$  & 0.2550 & 0.2364  &   0.2146 & 0.2046 &0.2027 
&0.2049
\\
   $-1$  & 0.2079 & 0.1869   &  0.1740  &   0.1714
&0.1731  
&0.1740
\\
   $-10$  & 0.1897 &  0.1731  &  0.1670 & 0.1665 &0.1686
&0.1693
\\
   $-100$  & 0.1875 & 0.1714  &   0.1662 &  0.1659 
&0.1682 
&0.1688
\\
\hline
\end{tabular}
\end{center}
\end{table}
\begin{table}[!ht]
\begin{center}
\caption{Same as Table \ref{table1} for the chemical potential $\mu_0$.}
\label{table2}
\begin{tabular}{|r|r|r|r|r|r|r|}
\hline
$a\backslash N=$ & 10  & $10^2$  &  
$10^4$ & $10^5$ &$10^5\{4\}$ &   $10^5\{10\}$ \\
\hline
   $-0.001$  & 0.3449   &  0.3361  & 0.3333  &   0.3331
&0.3284  &
0.3312    \\
   $-0.01$  &  0.3443     & 0.3351   &   0.3304   &   0.3281 
& 0.3237  &
0.3263
\\
   $-0.1$  &  0.3310   & 0.3139    & 0.2880     &   0.2749  &  
0.2743
&
0.2753\\
   $-1$  &   0.2670    &  0.2474      &    0.2326     &   
 0.2290
&0.2315 
& 0.2314 \\
   $-10$  & 0.2409      &    0.2277     &  0.2225      &   
 0.2219 &0.2249 
& 0.2246\\
   $-100$  & 0.2376      &   0.2254     &   0.2215     &  
 0.2212
& 
0.2242 
&0.2241\\
\hline
\end{tabular}
\end{center}
\end{table}

\begin{figure}[tbp]
\begin{center}
{\includegraphics[width=\linewidth]{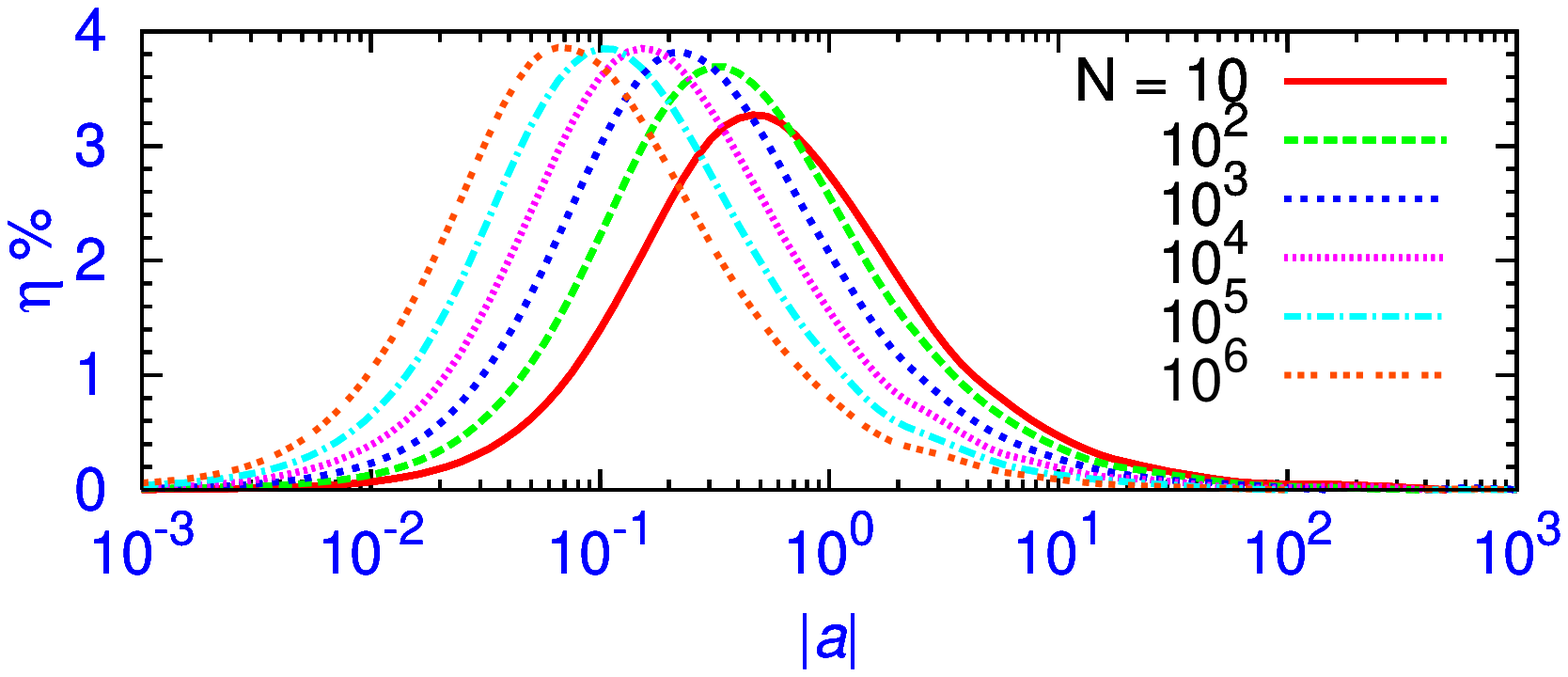}}
{\includegraphics[width=\linewidth]{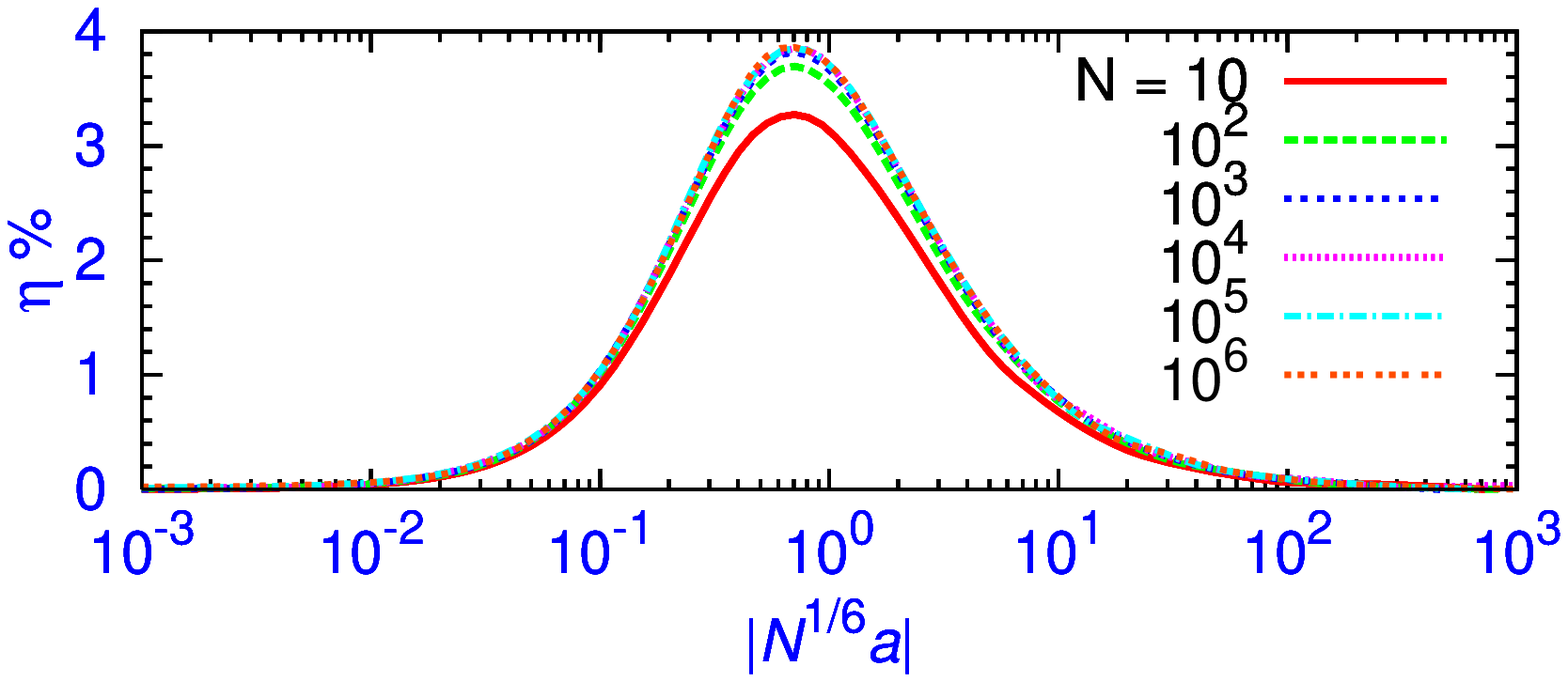}}
\end{center}
\caption{(Color online) (a) The plot of
$\eta = 100[E(N,a)-2\langle E_{\mathrm{pot}}\rangle ]/E(N,a)$ 
 for different $N$ vs.    $|a|$; and   (b) $\eta = 
100[E(N,\alpha)-2\langle E_{\mathrm{pot}}\rangle]/E(N,\alpha)$  vs.  
$|\alpha|$, 
$\alpha \equiv 
N^{1/6}a$. Scattering lengths are expressed in oscillator units.
}
\label{fig2}
\end{figure}

To quantify the deviation of the Fermi superfluid in 
the 
crossover region from universality, we study the deviation of 
our result for the 
trapped Fermi superfluid from the virial theorem, $E=2\langle  
E_{\mathrm{pot}} \rangle.$ For this purpose 
we plot the function $\eta=100 (E-2\langle 
E_{\mathrm{pot}}\rangle)/E$ vs. $|a|$ for different $N$ in 
Fig. \ref{fig2} (a). The shift of the curves with $N$ clearly 
shows that the results are dependent on  $a$ and 
$N$. We find that the maximum deviation from the virial theorem in the 
crossover region is quite small and is about $4\%$. 
A careful examination of Fig. \ref{fig2} (a) shows that this 
dependence 
on $|a|$  is linear in log scale and can be included by plotting 
$\eta$ vs. $|\alpha|, \alpha=N^{1/6}a$, as can be seen 
from Fig.    
\ref{fig2} (b). In  Fig. \ref{fig2} 
(b) we find that all  curves for 
$\eta$ have collapsed essentially on a single curve.
However, the  curve for $N=10$ is a bit different from others, which 
confirms  that there is a deviation from  universality for 
smaller $N$.   
This $|\alpha|=|N^{1/6}a|$ dependence of properties  of a trapped Fermi 
superfluid  is quite 
universal as we  see in the following. 

\begin{figure}[tbp] \begin{center}
{\includegraphics[width=\linewidth]{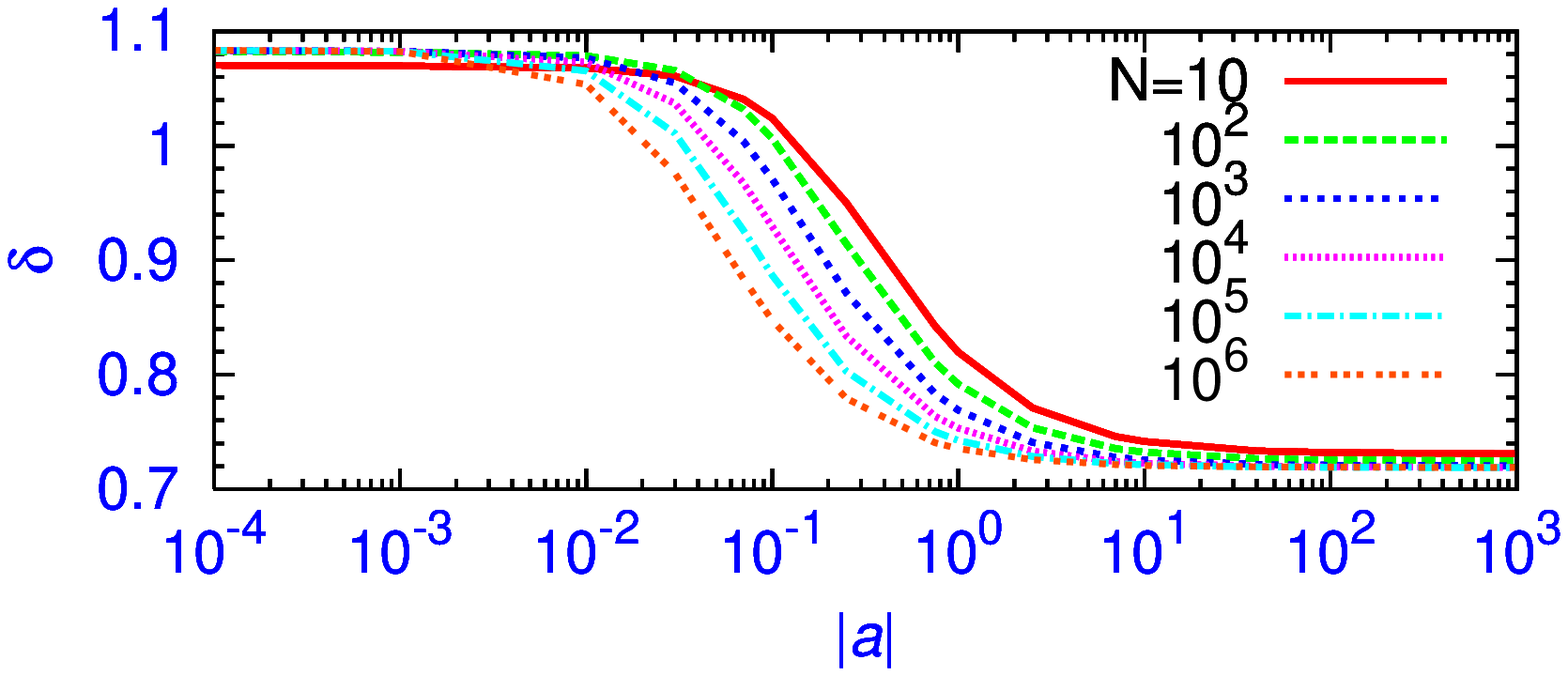}}
{\includegraphics[width=\linewidth]{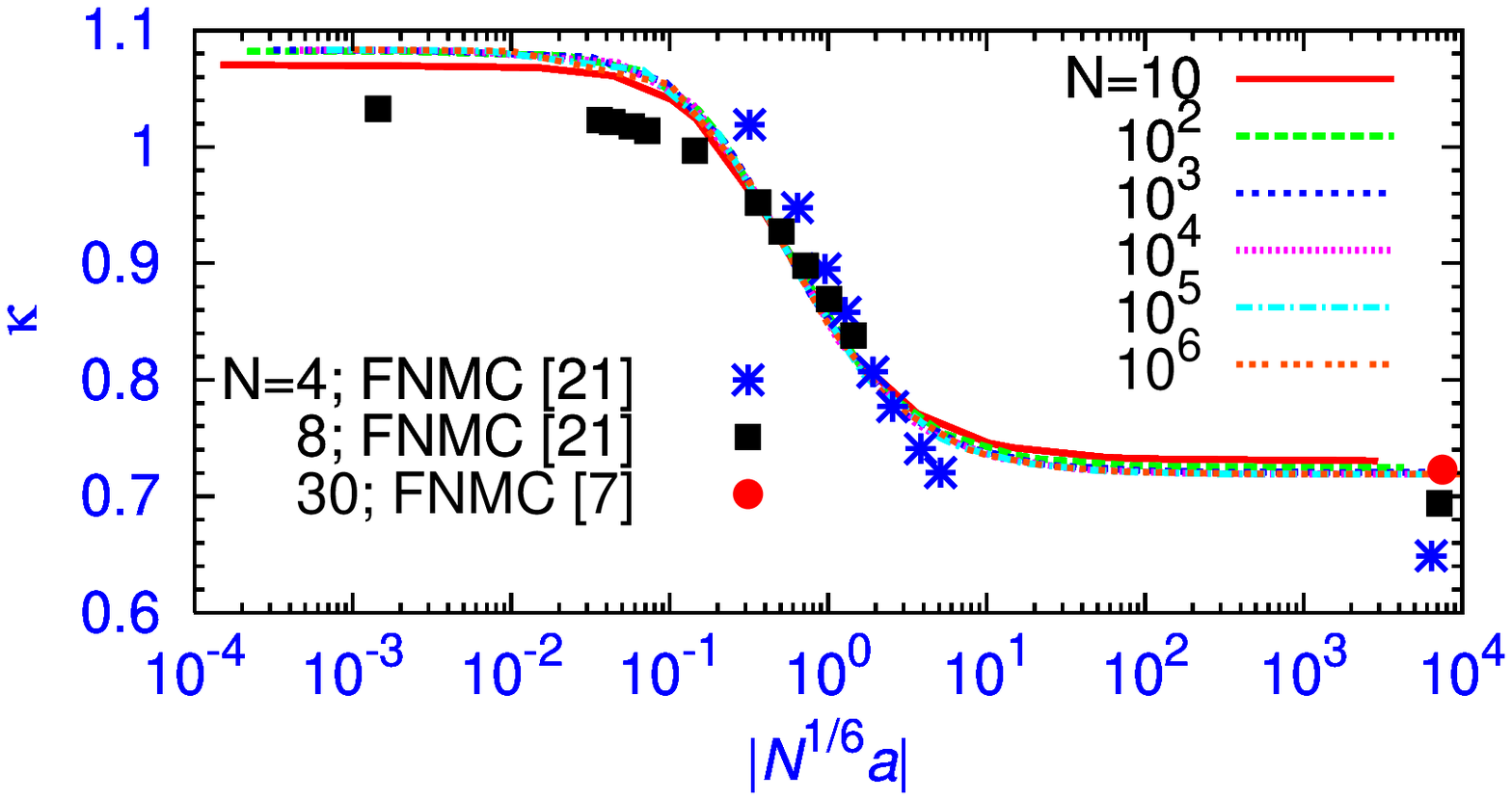}} 
{\includegraphics[width=\linewidth]{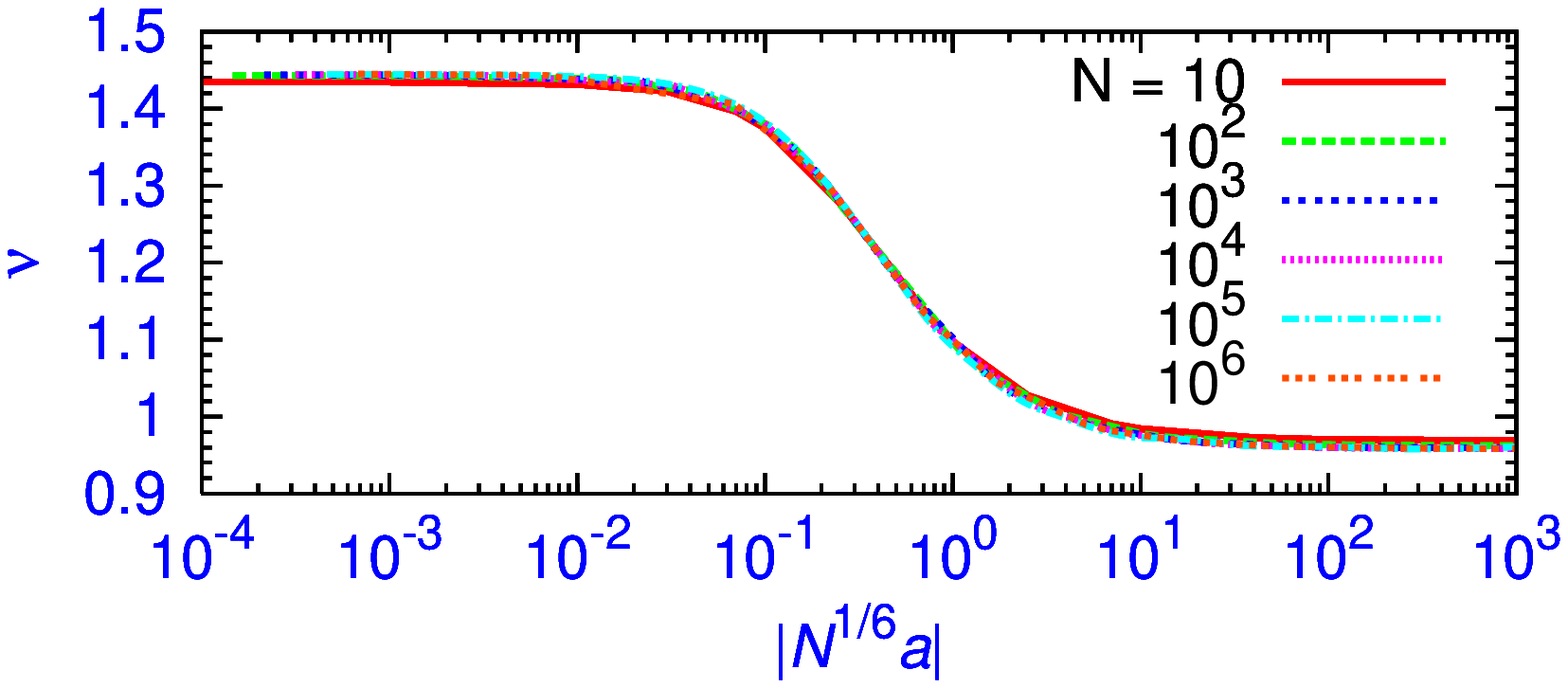}} 
\end{center}
\caption{(Color online) (a) 
$\delta\equiv [E(N,a)/N^{2/3}-0.37]/N^{2/3}$
vs $|a|$   for different $N$ for a
trapped
Fermi
gas from Eq. (\ref{1}).
(b) 
$\kappa\equiv [E(N,\alpha)/N^{2/3}-0.37]/N^{2/3}$ 
vs $|\alpha|$ for different $N$ 
from Eq. (\ref{1}) and FNMC method \cite{vs}.  (c) 
$\nu \equiv [\mu_0(N,\alpha)/N^{2/3}-0.27]/N^{2/3}$ vs
$|\alpha|$  for different $N$. (Energies and scattering lengths are expressed in oscillator units.)
}
\label{fig3}
\end{figure}

Now we study the universality in $E$ and 
$\mu_0$ of the trapped Fermi superfluid obtained from a 
solution of Eq. (\ref{1}). We plot the function 
$\delta \equiv [E(N,a)/N^{2/3}-0.37]/N^{2/3}$ vs. $|a|$   
for different $N$  in Fig. \ref{fig3} (a) 
[We recall  that the universal nature of the function $\delta$ was 
obtained 
from an analysis of results in Fig. \ref{fig1} (b)].  
We find that as in Fig. \ref{fig2} (a) the results for different $N$ are 
distinct. If the system were really dominated by universality, a plot of 
the scaled quantities $\kappa \equiv [E(N,\alpha)/N^{2/3}-0.37]/N^{2/3}$ 
vs. 
$|\alpha|$ would lead to universal curves. This is indeed found in Fig.  
 \ref{fig3} (b), where we also included the  
 FNMC results \cite{blume} for small $N=4,8$. For small $N$ 
there 
is some deviation from universality which disappears for $N>10$. We 
note that in Fig. \ref{fig3} (b)  the 
$N=30$  and in Fig. \ref{fig1} (b) the large-$N$ FNMC and GFMC data 
lie on the universal curves.  These 
universal curves yield the   simple formula for the energy of a 
large system
with $N$ atoms in terms of that of a small system with $N'$ 
atoms $(N>>N')$
\beqa\label{7}
E(N,\alpha)=N ^{4/3} \frac {E(N',\alpha)/N'^{2/3}-\gamma}{N'^{2/3}},
\eeqa 
with $\gamma=0.37$, and 
 where we have neglected the small constant 0.37 compared to the large 
quantity $E(N,\alpha)/N^{2/3}$. When $N$ and $N'$ are both large, 
formula  
(\ref{7}) becomes $E(N,\alpha)=E(N',\alpha)(N/N')^{4/3}$. 
We note that, in Eq. (\ref{7}), energies are 
to be considered for the same $\alpha$ and not $a$.
We indeed calculated the energies 
for $N=10^5$ atoms using the data for $N'=4$ and 10. The 
predictions so 
obtained for $N'=4,10$, listed Table \ref{table1}, compare well 
with the calculated results within about  $3\%$  and  $2\%$ errors, 
respectively.  In 
Fig. 
\ref{fig3} 
(c) we plotted scaled chemical potentials vs. $|\alpha|$ 
 for $N=10$ to $10^6$ and find that they all lie 
on the same universal curve. In this case also a prediction of chemical 
potential for large $N$ using the same for a small $N'$ can be made 
through the scaling formula (\ref{7}) but now with $\gamma=0.27 $ as 
obtained from an analysis of  Fig. \ref{fig1} (b). 
The predicted $\mu_0$ for $N=10^5$  using the data for 
$N'=4$ and 10, listed in Table \ref{table2}, 
compare 
well with the calculated results within an error of less than $1.5\%$. 

\begin{figure}[tbp] \begin{center}
{\includegraphics[width=\linewidth]{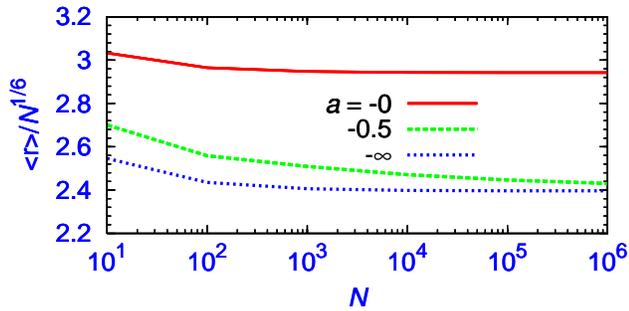}} 
\end{center}
\caption{(Color online) $\langle r \rangle /N^{1/6}$  of a trapped Fermi 
superfluid vs. $N$ for different $a=-0, -0.5,-\infty$. (Lengths are expressed in oscillator units.)
}
\label{fig4}
\end{figure}

Because of the universal behavior due to the dominance of Pauli 
repulsion,
other observables  of a Fermi superfluid are also correlated with 
energy 
and atom number. For example, the root mean square
radius $\langle r \rangle$ has the scaling  $\langle r \rangle \sim 
N^{1/6}$  in BCS and unitarity limits.
In Fig. \ref{fig4} we plot 
 $\langle r \rangle/N^{1/6}$ vs. $N$ 
for different $a$.  Again 
there is a slight violation of this scaling for small $N$ and 
in the   crossover region, which could be remedied  by a fine tuning
of the type (\ref{7}). Nevertheless, this clearly 
shows 
that the radius of a large trapped Fermi superfluid is predictable from 
the  knowledge of radius of a small system not only in the BCS and 
unitarity limits but also along the crossover.  In the case of nuclei, 
where Pauli repulsion plays an important role, correlations exist among 
binding energy, 
radius and the number of nucleons \cite{bw}.

\section{Conclusion}

In conclusion, from an extensive numerical study of the static 
properties of a 
trapped two-component Fermi superfluid using a Galilei-invariant DF 
formulation \cite{LS}, equivalent to a generalized hydrodynamic 
formulation with 
the correct phase-velocity relation \cite{review},  we establish that, 
because of the 
dominance of the Pauli repulsion, the trapped Fermi superfluid has a 
universal behavior not only in the BCS and unitarity limits but also in 
the crossover region. This allows a prediction of the static 
properties  (energy, chemical potential, rms radius etc.) of a large 
Fermi 
superfluid in the crossover region from a 
knowledge of the same of a small system through a  scaling 
relation, cf. Eq. (\ref{7}). 
The thus predicted 
energy and chemical potential of a system with $10^5$ atoms from a 
knowledge of the same with 4   (10) atoms is found to have an 
error of less than 3$\%$  (2$\%$). Actually, for small systems one needs 
to introduce finite-size effects, such as, higher-order gradient 
corrections and shell effects. The $2-3 \%$ discrepancy quantifies the 
contribution of such effects. 
Although we used a
DF formulation in the present study,  
due to the dominance of Pauli repulsion and existence of robust 
scalings, we do not believe our conclusion  to be so peculiar as to have 
no 
general validity.

FAPESP
and CNPq (Brazil) provided partial support.
Research was (partially) completed
while S.K.A. was visiting the Institute for Mathematical Sciences,
National University of Singapore in 2007.


\end{document}